\documentclass[prl,twocolumn,byrevtex,twocoloumn,byrevtex,floatfix,superscriptaddress]{revtex4}

\usepackage{amssymb,amsmath,graphicx}
\begin{document}

\title{Limits of Validity of Rashba Model in BiTeI: A High-field Magneto-optical Study}

\author{S. Bord\'acs}
\affiliation{Department of Physics, Budapest University of
Technology and Economics, 1111 Budapest, Hungary}
\affiliation{Hungarian Academy of Sciences, Premium Postdoctor
Program, 1051 Budapest, Hungary}

\author{M. Orlita}
\affiliation{Laboratoire National des Champs Magn\'etiques Intenses,
CNRS-UGA-UPS-INSA-EMFL, 25, avenue des Martyrs, 38042 Grenoble, France}
\affiliation{Institute of Physics, Charles University, Ke Karlovu 5, 12116 Praha 2, Czech Republic}

\author{M. \v{S}ikula}
\affiliation{Laboratoire National des Champs Magn\'etiques Intenses,
CNRS-UGA-UPS-INSA, 25, avenue des Martyrs, 38042 Grenoble, France}

\author{H. Murakawa}
\affiliation{Department of Physics, Osaka University, Toyonaka 560-0043, Japan}

\author{Y. Tokura}
\affiliation{RIKEN Center for Emergent Matter Science (CEMS), Wako, Saitama 351-0198, Japan}
\affiliation{Department of Applied Physics, University of Tokyo, Hongo, Tokyo 113-8656, Japan}

\date{\today}

\begin{abstract}
It was recently shown that BiTeI, a semiconductor with polar crystal structure, possesses a giant spin-splitting of electrons, which has been interpreted in terms of Rashba-type spin-orbit coupling. Here, we use high field magneto-optical spectroscopy to quantify the deviations of the conduction-band profile from this appealing, but at the same time, strongly simplifying model. We find that the optical response -- comprising a series of inter-Landau level excitations -- can be described by the Rashba model only at low magnetic fields. In contrast, the high-field response appears to be more consistent with a simple picture of massless electrons in a conical band. This points towards more linear rather than parabolic at energies well above the bottom of the conduction band.  
\end{abstract}

\maketitle

The interplay of broken inversion symmetry and spin-orbit interaction gives rise to new states of matter such as the helical surface state of topological insulators \cite{Hsieh2008, Xia2009} or the skyrmion spin texture in ultra-thin films \cite{Heinze2011} and in non-centrosymmetric bulk magnets \cite{Muhlbauer2009, Yu2010}. These systems attract much attention owing to the numerous intriguing phenomena which has been observed or predicted in them, e.g. spin-, topological- and quantum anomalous Hall-effects \cite{Murakami2003,Neubauer2009,Chang2013,Checkelsky2014}, or topological superconductivity with Majorana edge modes \cite{Hell2017, Pientka2017}. Moreover, these fundamentally new states may find applications in spintronics or in topological quantum computation \cite{Pesin2012,Bravyi2002}.

\begin{figure}[t!]
\includegraphics[width=\columnwidth]{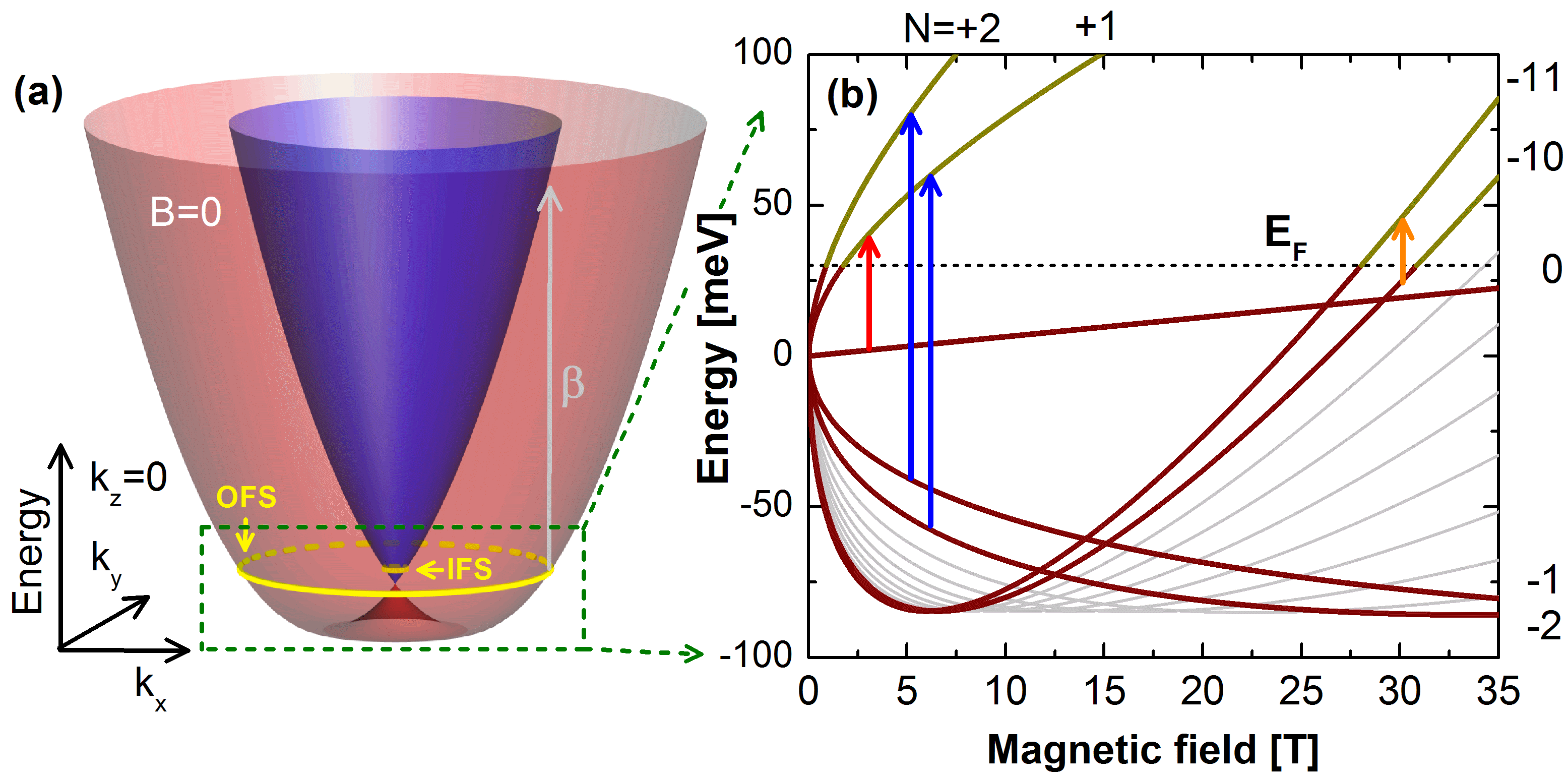}
\caption{(Color online) (a) A schematic view of energy-momentum dispersion relation implied by the Rashba model, with the inner and outer Fermi surfaces indicated by yellow lines (IFS and OFS, respectively). (b) The LLs spectrum implied by the Rashba model for $m$=0.09\,$m_0$ and $v$=5.75$\times$10$^5$\,m/s.  Vertical arrows indicate the electric dipole transitions between the filled (dark brown) and empty states (dark yellow).}
\label{fig1}
\end{figure}

One of the most fundamental model describing itinerant electrons with spin-orbit interaction in the lack of inversion symmetry was developed by Rashba \cite{Rashba1959}. Beside the kinetic energy, the model contains a spin-momentum coupling term linear both in momentum $\textbf{p}$ and in the spin of the electron, which is allowed by a polar field assumed to be oriented along the z-axis, $\mathbf{\hat{z}}$:
\begin{equation}
\mathcal{H}_R=\frac{\mathbf{p}^2}{2m}+\frac{\alpha}{\hbar}{\bf\hat{z}}\cdot({\boldsymbol\sigma}\times\mathbf{p}),
\label{eqRashba}
\end{equation}
where $m$ is the effective mass of the electron, $\boldsymbol{\sigma}$ are the Pauli matrices and $\alpha$ is the Rashba parameter. Due to the spin-orbit interaction the double degeneracy of the parabolic band is lifted: $\varepsilon_\pm$(k)=$\hbar^2$k$^2$/2$m$ $\pm\alpha$k (see Fig.~\ref{fig1}), and the electron spin whirls clockwise or counterclockwise around the center of the Brillouin zone in the k-space. This simple model was first used to describe the bulk band structure of semiconductors with polar wurtzite structure such as CdS and CdSe \cite{Rashba1959}, and later applied to two dimensional electron gases subject to structure inversion asymmetry \cite{Bychkov1984} or to surface states of heavy metals \cite{LaShell1996}.

So far, the largest Rashba parameter was found in the polar semiconductor BiTeI, which has Bi layers situated asymmetrically in between a Te and an I layer. Spin- and angle-resolved photoemission spectroscopy (sARPES) concluded that electron states at or close to the surface have Rashba-like spin-split dispersion with $\alpha$=3.85\,eV\AA~\cite{Ishizaka2011} consistent with ab-initio calculations \cite{Bahramy2011}. According to a recent study \cite{Fulop2018} even a single layer of BiTeI can be stabilized on gold surface, in which the coupling constant $\alpha$ is expected to be reduced to 2.1\,eV\AA. The band structure of bulk BiTeI has been studied by Shubnikov-de Haas (SdH) oscillations \cite{Martin2013,Bell2013,Murakawa2013,Ye2015} and optical spectroscopy \cite{Lee2011,Demko2012}, which also indicate the existence of two Fermi surfaces corresponding to the spin split inner and outer Fermi surfaces, IFS and OFS, respectively.

The majority of magneto-transport and magneto-optical experiments performed so far on BiTeI, have been interpreted on the basis of the Landau level (LL) spectrum implied by the Rashba model:
\begin{align}
\varepsilon_N^\pm &= \hbar\omega_cN\pm\sqrt{\left(\frac{\hbar\omega_c}{2}-\frac{g\mu_B}{2}B\right)^2+2e\hbar v^2NB}, \nonumber \\
\varepsilon_0 &= \frac{\hbar\omega_c}{2}-\frac{g\mu_B}{2}B,
\label{eqLLspectrum}
\end{align}
where $B$ is the external magnetic field applied along the $\mathbf{\hat{z}}$  direction, $\omega_c$=$\frac{eB}{m}$ is the cyclotron frequency related to the quadratic part of the dispersion $\varepsilon_\pm$(k), $v$=$\frac{\alpha}{\hbar}$ is the velocity parameter at the crossing point of the spin-polarized parabolic bands, $g$ is the spin only g-factor, and $N$ is a positive integer \cite{Rashba1959}. As shown in Fig.~\ref{fig1}, there are two series of LLs corresponding to the two spin-split bands: $\varepsilon_N^+$ monotonously increases with field while $\varepsilon_N^-$ decreases in low fields till it reaches the bottom of the $\varepsilon_-$ band and then it also increases. The two different SdH oscillation frequencies have been assigned to the series of LLs crosses the IFS and OFS as the field is increased \cite{Martin2013,Bell2013,Murakawa2013,Ye2015}. 

On the other hand the results of the previous low magnetic field cyclotron resonance study can be explained by a single conical band \cite{Bordacs2013}. Since the Rashba energy dominates the dispersion, when the Fermi-energy is close to the band crossing point, the energy levels and correspondingly the observed transition energies follow square root dependence both on the LL index, $N$ and the magnetic field: $\varepsilon_N^\pm\approx$v$\sqrt{2e\hbar NB}$, which is characteristic of Dirac fermions. Furthermore, the selection rules are also identical in the two cases. The electric dipole term excites electrons from state $N$ to states $N\pm1$ irrespective of the $\pm$ index of the initial or final states.

In higher magnetic fields the LL spectrum implied by the Rashba model significantly deviates from the one known for massless Dirac electrons. The doubly degenerate transition of the conical model are split as $\varepsilon$(0$\rightarrow$1$^+$)-$\varepsilon$(1$^-\rightarrow$0)=$\hbar\omega_c$+g$\mu_B$B and $\varepsilon$(N$^-$$\rightarrow$(N+1)$^+$)-$\varepsilon$((N+1)$^-$$\rightarrow$N$^+$)=2$\hbar\omega_c$ for N$\neq$0 due to the parabolic term in the dispersion $\varepsilon_\pm$(k). Such deviations from the conical band model are in principle observable in magneto-optical experiments, provided the cyclotron energy $\hbar\omega$ becomes larger, or at least, comparable with the width of inter-LL resonances.

In this paper, we study the bulk band structure of BiTeI using high-field LL spectroscopy, which provides us with relatively high spectral resolution. In contrast to our former study \cite{Bordacs2013}, we measured directly the field induced changes in the absorption spectrum by detecting the light transmission through thin flake samples. We show that the optical response due to inter-LL excitations in BiTeI can be described in a broad range of applied magnetic fields (up to 34\,T) using a simple Dirac-type model for massless electrons in a conical band. This observation limits the quantitative validity of the Rashba model to a relatively narrow range of momenta around the band crossing point, where the Rashba and Dirac-type models imply nearly the same magneto-optical response. 

Using a Leica microtome thin $ab$ plane cuts were prepared from a single crystal of BiTeI which was grown by the Bridgman method as described in Ref.~\onlinecite{Ishizaka2011}. Only the thinnest slices, which had a thickness of approximately 1-2\,$\mu$m, were transparent enough for the transmission measurements. Infrared absorption spectra were measured in the High Magnetic Field Laboratory Grenoble (LNCMI-G) using a commercial Bruker Fourier-transform spectrometer. The radiation from the spectrometer, which is guided by a light-pipe, is transmitted through the sample and detected by a bolometer placed directly below the sample. The temperature of the sample was 2\,K during the measurement, whereas magnetic fields up 13\,T and 34\,T were provided by a superconducting solenoid and by a resistive coil, respectively.

\begin{figure}[t!]
\includegraphics[width=3 in]{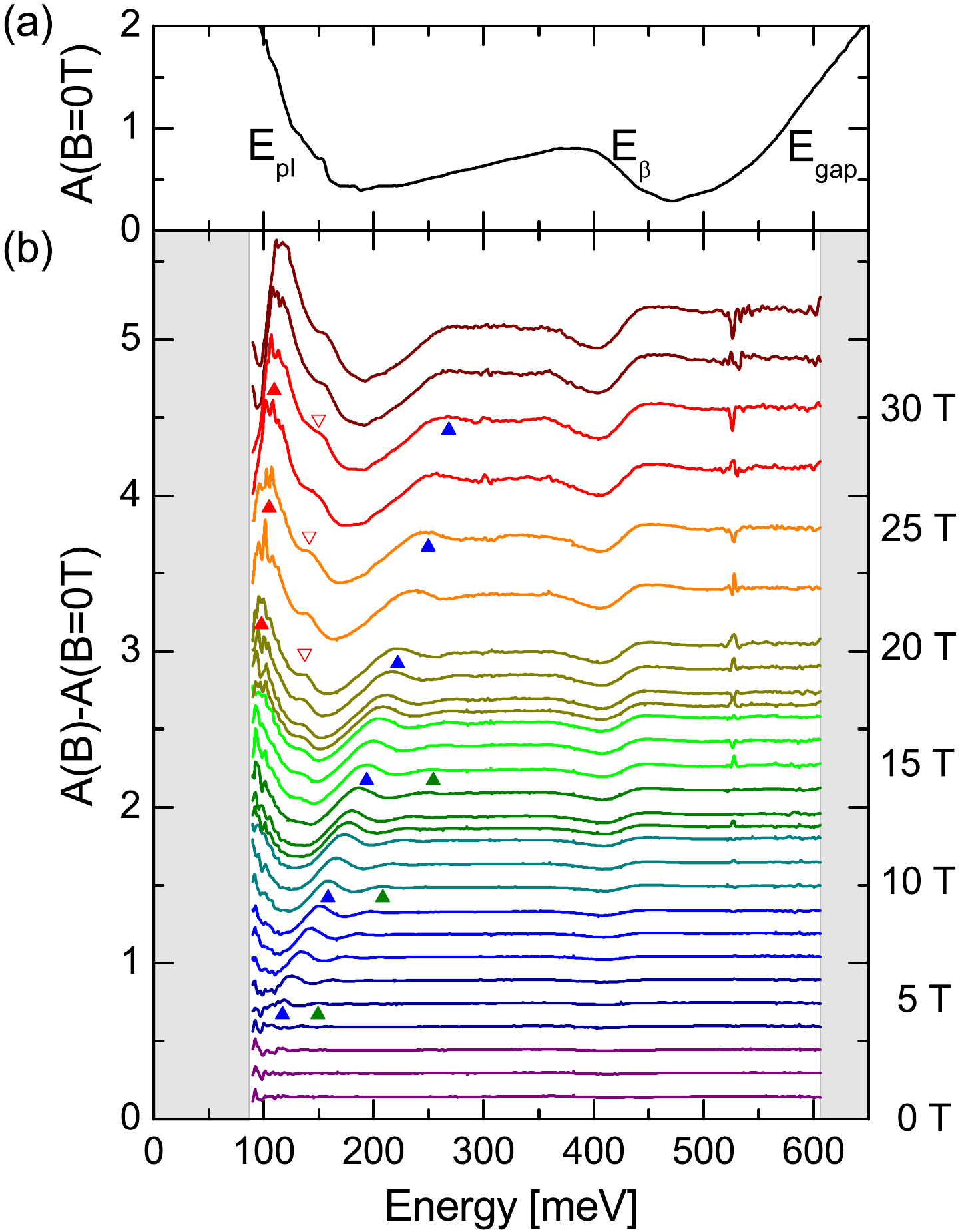}
\caption{(Color online) (a) Zero-field absorbance spectrum, which is derived from the transmission, $T$ as $A$=-log($T$). $E_{pl}$, $E_\beta$ and $E_{gap}$ are the energy of the plasma edge, $\beta$-transition and the band gap, respectively. (b) Magnetic field dependence of the relative absorbance spectra. For clarity the spectra are shifted in proportion with the magnetic field. The tip of the solid arrows show the position of the inter Landau-level transitions as plotted in Fig.~\ref{fig3}. The empty arrows marks an unresolved absorption peak at the high energy side of the 0$\rightarrow$1 transition.}
\label{fig2}
\end{figure}

A typical zero-field absorbance spectrum, $A$ is shown in Fig.~\ref{fig2} (a), which is derived from the measured transmission, $T$ as $A$=-log($T$). Since the absolute value of the intensity could not be measured the scale of the absorbance is arbitrary. At low photon energies, the transmission drops significantly below the plasma edge ($E_{pl}\approx 125$\,meV). At high photon energies, the transmission window closes due to interband excitations across the fundamental energy band gap ($E_g \approx 600$\,meV). The relatively abrupt increase of the absorption at photon energies below $E_{\beta}\approx 400$\,meV is due to the onset of excitations between spin-split conduction band, see the $\beta$ line in Fig.~\ref{fig1} (a).

The collected magneto-absorbance spectra are presented in Fig.~\ref{fig2} (b), always normalized by the zero-field absorbance and corrected for the field-induced variation of the bolometer's response. With increasing field a series of maxima appear in the relative magneto-absorbance spectra $A(B)$-$A(B=0)$, which is associated with individual inter-LL excitations, see Fig.~\ref{fig1} (b). An additional modulation appears around the photon energy of $E_{\beta}$. This modulation reflects the magnetic field induced splitting of high-energy onset of absorption between spin-split conduction band ($\beta$ transition).

\begin{figure}[t!]
\includegraphics[width=\columnwidth]{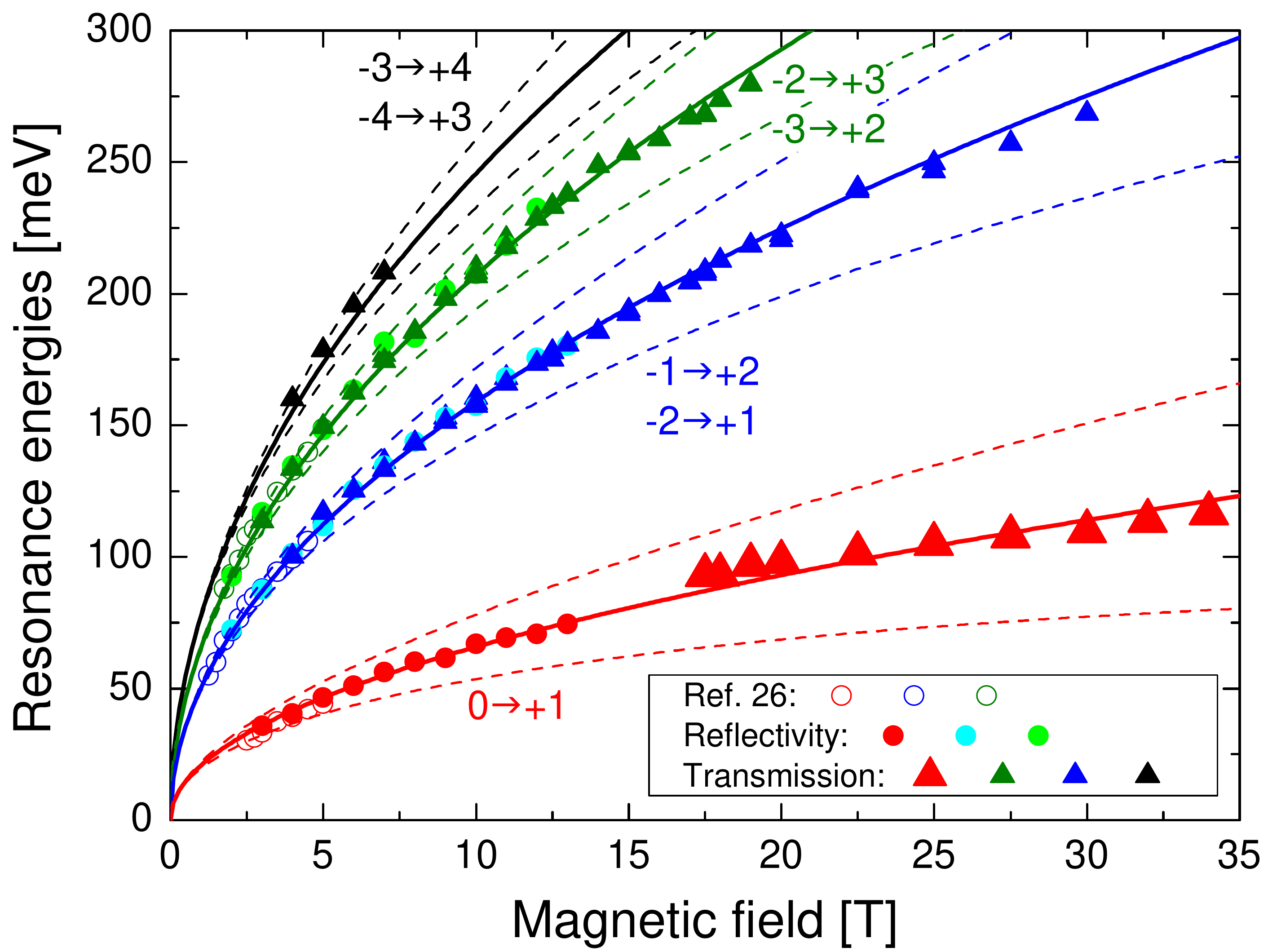}
\caption{(Color online) The magnetic field dependence of the inter Landau-level transitions. Filled trianguls (circles) indicate the experimental resonance energies obtained from measurements in transmission (reflection) geometry. Open circles are reproduced from Ref.~\onlinecite{Bordacs2013}. Solid lines represent a fit with a conical model which gives $v$=5.75$\times$10$^5$\,m/s. Dashed lines show the Landau-level dispersion in the Rashba model with the same $\alpha$=$v\hbar$ and $m$=0.09\,$m_0$, $g$=2.}
\label{fig3}
\end{figure}

The field-dependence of inter-LL resonances are plotted in Fig.~\ref{fig3}. These data are complemented with positions of resonances observed in additional magneto-reflectivity experiment performed on a sample from the same batch (up to 13\,T) and compared with results from Ref.~\onlinecite{Bordacs2013}. Importantly, the positions of all resonances observed in this study, but also those from Ref.~\onlinecite{Bordacs2013}, may be fitted using a simple model that assumes electric-dipole excitations ($N$$\rightarrow$$N\pm1$) between Landau levels of  electrons with conical dispersion. Within such a single-cone model, the field dependence of the resonances can be fitted with a series of square root of $B$ curves, using the slope of the conical band $v$ as the only fitting parameter. The resulting fit describes the experimental data rather well and implies $v$=5.75$\times$10$^5$\,m/s.

Importantly, this conclusion is clearly not consistent with expectation based on the Rashba model, in which absorption line splits into two components with the increasing magnetic field due to the parabolic (kinetic) term in Eq.~\ref{eqRashba}. The expected positions of inter-LL resonances within the Rashba model are plotted with the dashed lines in Fig.~\ref{fig3}, which were calculated using the full LL spectrum in Eq.~\ref{eqLLspectrum} for $m$=0.09\,$m_0$ and $\alpha$=$v\hbar$=3.785\,eV\AA. In our experimental data, we find no traces of such splitting. This indicates that the linear dispersion is a better approximation (rather than quadratic dispersion) in fairly broad range of energies around the band crossing point at $k$=0. The Zeeman term with non-zero g-factor would result in a deviation from the $\sqrt{B}$ field dependence of the transition energy of the 0$\rightarrow$1$^+$ and the 1$^-$$\rightarrow$0 transitions and the corresponding splitting should appear in the magneto-absorbance spectra. Within the accuracy of the measurement none of these effects are detected, thus, the energy of the 0th LL is field independent, and the g factor is negligible.

Interestingly, a high-energy shoulder develops on the 0$\rightarrow$1 transition above $B$$>$10\,T and its position weakly increases with the magnetic field (open triangles in Fig.~\ref{fig2}). Its position in the spectrum rather well coincides with the plasma energy $E_{pl}$$\sim$120\,meV. However, we do not see any apparent mechanism which would enable coupling of the longitudinal plasmon wave with the transversal optical wave in the present experimental configuration. 
The high-energy shoulder of the 0$\rightarrow$1 absorption line, or in general the line asymmetry, may be in a bulk material related to the particular profile of the joint density of states, which reflects different c-axis dispersion of electrons in the $n=0$ and $n=1$ LLs. However, this effect can only cause a high energy tail in the joint density of states decaying as $1/\sqrt{E}$, which cannot explain the observed side-peak. Another explanation could be that the high-energy shoulder may appear due to the splitting of 0$\rightarrow$1$^+$ and 1$^-$$\rightarrow$0 transitions, but the overall character of the high-energy shoulder -- the magnetic field dependece of the position and intensity, in particular -- do not make this option probable.

Let us now discuss the modulation of $A(B)$-$A(B=0)$ spectra which appears around the high-energy onset of absorption between spin-split conduction band (around transition $\beta$ in Fig.~\ref{fig1} (a)). This modulation may be straightforwardly explained in terms of inter-LL excitations, when electrons are promoted by incoming radiation from the highest occupied N$^-$ LL in the lower spin-split (OFS) conduction band to $(N-1)$$^+$ and $(N+1)$$^+$ level in the upper band. The absorption edge at $E_\beta$ then becomes split by the energy of $2\hbar\omega_c$ in the Rashba model. However, in BiTeI the energy of the splitting cannot be resolved due to line broadening, thus, it translates in the relative magneto-transmission spectra $A(B)$-$A(B=0)$ into a horizontal-s-like profile around the energy of $E_\beta$=430\,meV.  

A simple Rashba model predicts the energy of $\beta$ transition to be $\sim$700\,meV for the above deduced value of the velocity parameter, $\alpha$= $v\hbar$=3.785\,eV\AA and $m$=0.09\,$m_0$. This discrepancy, that the Rashba model fails to consistently describe the position of the $\beta$ transition, can also be noticed if its carrier density dependence is analysed as in a previous MOKE study \cite{Demko2012}. The energy of the $\beta$ transition almost independent of the carrier density, which is also rather consistent with linear bands in the conduction band.

In this paper, we report the observation of a series of inter-LL transitions in a BiTeI sample placed in high magnetic fields. All of the transitions can be explained by a conical band with a velocity parameter $v$=5.75$\times$10$^5$\,m/s. Even though the conduction band of BiTeI is nowadays routinely described using the Rashba Hamiltonian, we conclude that its validity is only qualitative in a broader range of momenta.

\section*{Acknowledgement}

We are grateful to J. G. Checkelsky, L. Ye, I. K\'ezsm\'arki and M. Potemski for fruitful discussions. We acknowledge the support of LNCMI-CNRS, a member of the European Magnetic Field Laboratory (EMFL). M.O. acknowledges the support by ANR through the DIRAC3D project. This work was supported by the Funding Program for World-Leading Innovative R\&D on Science and Technology (FIRST Program), Japan, and by the National Research, Development and Innovation Office – NKFIH, PD 111756, by the BME-Nanonotechnology and Materials Science FIKP grant of EMMI (BME FIKP-NAT).

\end{document}